\documentclass[twocolumn,showpacs,preprintnumbers,amsmath,amssymb,aps]{revtex4-1}
\pdfoutput=1
\usepackage[pdftex]{graphicx}

\usepackage{dcolumn}% Align table columns on decimal point\right] 
\usepackage{bm}% bold math

\usepackage{color}
\usepackage{amstext}

\begin{document}

%%%%%%%%%%%%%%%%%%%%%%%%%%%%%%%%%%%%%%%%%%%%%%%%%%%%%%%%%%%%%%%%%%%%%%%%%%
%                               Title                                    %
%%%%%%%%%%%%%%%%%%%%%%%%%%%%%%%%%%%%%%%%%%%%%%%%%%%%%%%%%%%%%%%%%%%%%%%%%%

\title{Frustrated multiband superconductivity}
\author{R. G. Dias}
\author{A. M. Marques}
%\email{rdias@fis.ua.pt}
\affiliation{Departamento de F\'{\i}sica, I3N, Universidade de Aveiro, Campus Universit\'{a}rio de Santiago,
\\
3810-193 Aveiro, Portugal.}
\date{\today}

%%%%%%%%%%%%%%%%%%%%%%%%%%%%%%%%%%%%%%%%%%%%%%%%%%%%%%%%%%%%%%%%%%%%%%%%%%
%                              abstract                                  %
%%%%%%%%%%%%%%%%%%%%%%%%%%%%%%%%%%%%%%%%%%%%%%%%%%%%%%%%%%%%%%%%%%%%%%%%%%

\begin{abstract}
We show that  a clean
multiband superconductor may display one or several phase transitions with increasing temperature  from or to frustrated  configurations of the relative phases of the superconducting order parameters. These transitions may occur when  more than two  bands are involved in the formation of the superconducting phase and when the number of repulsive  interband interactions is odd.  These transitions are signalled by slope changes in the temperature dependence of the superconducting gaps.
\end{abstract}


\maketitle
%%%%%%%%%%%%%%%%%%%%%%%%%%%%%%%%%%%%%%%%%%%%%%%%%%%%%%%%%%%%%%%%%%%%%%%%%%
%                              body of paper                             %
%%%%%%%%%%%%%%%%%%%%%%%%%%%%%%%%%%%%%%%%%%%%%%%%%%%%%%%%%%%%%%%%%%%%%%%%%%
%\section{Introduction}
The possibility of multiple bands contributing to the formation of a superconducting phase has  been considered in the case of transition metals \cite{Garland1963,Kondo1963,Suhl1959},  superconducting copper oxides \cite{Combescot1995}  and magnesium diboride \cite{Gonnelli2002,Iavarone2002,Tsuda2003}. More recently, sign-reversed  two-band superconductivity has been proposed for the  iron-based layered pnictides \cite{Kuroki2008,Mazin2008}. In the two-band case, the relative phase of the superconducting gap functions associated to each band   is  determined by the sign of the interband interaction, being zero  ($\pi$) for attractive (repulsive) interband coupling. However, if more than two electronic bands have to be considered in the study of the superconducting phase, the relative phases are not uniquely defined by the signs of the interband interactions and in particular, frustration may occur if the number of repulsive  interband interactions is odd.

There is a close analogy between such frustrated multiband superconductors and the well studied problem of  frustrated Josephson junction arrays  since the interband pairing may be regarded as an interband Josephson tunnelling \cite{Leggett1966}.  For example, it is known that a squared Josephson junction array with a $\pi$ magnetic flux per plaquette is frustrated with a degenerate ground state.  The effect of the $\pi$ magnetic flux is the change of the sign of the Josephson coupling from positive to negative, i. e., a $\pi$-junction \cite{Li2003133}. Such $\pi$-junctions are also present without magnetic flux as a consequence of d-wave pairing  symmetry. In the case of a frustrated (odd number) loop of $\pi$-junctions, a spontaneous current will be present. 

Studies of frustrated Josephson junction arrays assume usually symmetric junctions. The case of multiband superconductivity is more closely analogous to the case of an array of asymmetric Josephson junctions \cite{Anderson1964}. When temperature is increased, strong modifications of the ratios of the interband Josephson tunnelling rates may occur due to the  relative changes of the superconducting gaps which are known to happen in multiband superconductors \cite{Suhl1959}. We show in this manuscript that this may lead to one or two  phase transitions with increasing temperature  from or to frustrated   configurations  of the relative phases of the superconducting order parameters (which correspond to degenerate ground states which are chirally different). 
In very recent works, the ground state of a three-band superconductor with repulsive interband interactions has been studied using a phenomenological Ginzburg -Landau approach \cite{Tanaka2010} and a simplified BCS gap equation system where all  intraband interactions are zero \cite{Tanaka2010a}. The effects reported in this manuscript are not observed neither in the temperature range where the Ginzburg-Landau approach is valid (since the gap functions have constant ratios in this temperature range) nor when all  intraband interactions are zero. Furthermore, while the BCS gap equation does indicate the maxima, minima and saddle points of the free energy, it does not indicate which solution corresponds to the absolute minimum of the free energy. 

In this paper, we address the case of a three-band superconductor with one or more repulsive interband couplings. This study can easily be generalized to any number of superconducting bands.

%%%%%%%%%%%%%%%%%%%%%%%%%%%%%%%%%%%%%%%%%%%%%%%%%%%%%%%%%%%%%%%%%%%%%%%
%                              Hamiltonian                            %
%%%%%%%%%%%%%%%%%%%%%%%%%%%%%%%%%%%%%%%%%%%%%%%%%%%%%%%%%%%%%%%%%%%%%%%
%\section{Hamiltonian}
We adopt the Hamiltonian introduced by Suhl, Matthias, and Walker \cite{Suhl1959} to describe two-band superconductors, generalized for $n$ bands,
\begin{eqnarray}
        H &=& \sum_{i{\bm k} \sigma} \xi^i_{{\bm k} } 
        c^{i\dagger}_{{\bm k} \sigma} c^i_{{\bm k} \sigma}
        \\
        &-&  \sum_{ij,{\bm k}_1 {\bm k}_2} V_{ij}
        c^\dagger_{i,{\bm k}_1 \uparrow} c^\dagger_{i,-{\bm k}_1 \downarrow}
        c_{j,-{\bm k}_2 \downarrow} c_{j{\bm k}_2\uparrow},
        \nonumber
\end{eqnarray}
with 
$\xi_{\bm k}=\epsilon_{\bm k}-\mu$ and where $i$, $V_{ij}$,  and $\mu$ are respectively the band index ($i=1,\dots , n$), the interaction constant and the chemical potential. The ${\bm k}$ sums in the interaction term follow the usual BCS restrictions. S-wave symmetry is assumed  throughout the paper for simplicity.  
%The interband pairing may be regarded as an internal Josephson coupling.
The interaction constants are assumed to be positive (attractive interactions) for electrons in the same band ($i=j$), and one has $V_{ij}=V_{ji}$.

%%%%%%%%%%%%%%%%%%%%%%%%%%%%%%%%%%%%%%%%%%%%%%%%%%%%%%%%%%%%%%%%%%%%%%%%%
%                         Coupled gap equations                         %
%%%%%%%%%%%%%%%%%%%%%%%%%%%%%%%%%%%%%%%%%%%%%%%%%%%%%%%%%%%%%%%%%%%%%%%%%
%\section{Zero temperature}
In this paragraph, we follow the seminal discussion of  Ref.\cite{Leggett1966}. We start by restricting the $n$-band superconducting system to the subspace constructed from the  set of BCS states,
\begin{equation}
        \vert \mathcal{F} (\bm \Delta, \bm \Phi)\rangle =
        \prod_{j=1}^n  
        \left[
        \prod_{\bm k} 
        ( u_{j\bm k}+ e^{i \phi_j} v_{j\bm k} c^\dagger_{j {\bm k} \uparrow}
        c_{j \bm{-k} \downarrow} ) \right]
        \vert \Phi_0 \rangle ,
\end{equation}
where  $\bm \Delta=(\Delta_1,\cdots,\Delta_n)$ and $\bm \Phi=(\phi_1,\cdots,\phi_n)$ are the  absolute values of the superconducting gaps and the respective phases, and with the usual BCS definition of superconducting order parameters
\begin{equation}
        u_{i\bm k} v_{i\bm k}
        =
        \frac{\Delta_i}{2 \sqrt{[E^2_{i\bm k}+  \Delta_i^2]}} ,
\end{equation}
and $ u_{i\bm k}^2 + v_{i\bm k}^2 =1$ ($ u_{i\bm k}$ and $ v_{i\bm k}$ are real).  Such states with different values of $\Delta_i$ are orthogonal in the thermodynamic limit  since their inner product gives a Dirac $\delta$-function \cite{Leggett1966}. Defining the operators 
$
        \hat \Psi_i = \sum_{\bm k} c^\dagger_{i\bm k \uparrow}
         c^\dagger_{i\bm{-k} \downarrow}
$
which are diagonal in the previous basis in the thermodynamic limit, 
\begin{eqnarray}
        \hat \Psi_j \vert \mathcal{F} (\bm \Delta, \bm \Phi)\rangle
        &=&
        \Psi_j \vert \mathcal{F} (\bm \Delta, \bm \Phi)\rangle 
        \\
        &=&
        \sum_{\bm k} u_{j\bm k} v_{j\bm k}  e^{-i \phi_j}
        \vert \mathcal{F} (\bm \Delta, \bm \Phi)\rangle , 
        \nonumber
\end{eqnarray}
one may write a simple expression for the eigenvalues of the Hamiltonian which is also diagonal in this basis, 
\begin{eqnarray}
       E
      &=&
      \sum_i
      f_i(\left\vert \Psi_{i} \right\vert^{2})
      -
      \sum_i
      V_{ii} \left\vert \Psi_{i} \right\vert^{2} \nonumber \\
      &-&
      \sum_{i \neq j} 
      V_{i,j} \left\vert \Psi_{i} \right\vert
      \left\vert \Psi_{j} \right\vert \cos(\phi_{j}-\phi_{i}),  
      \label{eq:zeroTenergy}
\end{eqnarray}
where $f_i(\left\vert \Psi_{i} \right\vert^{2})$ is the kinetic energy contribution of the respective band term in the Hamiltonian and has an elaborate dependence on the superconducting parameters $\Delta_i$. The other terms result from the intraband and interband pairing terms in the Hamiltonian.
%%%%%%%%%%%%%%%%%%%%%
%      figure       %
%%%%%%%%%%%%%%%%%%%%%
\begin{figure}
	\begin{center}
		$\begin{array}{c@{\hspace{.1in}}c@{\hspace{.1in}}c}
		\includegraphics[width=2.1cm]{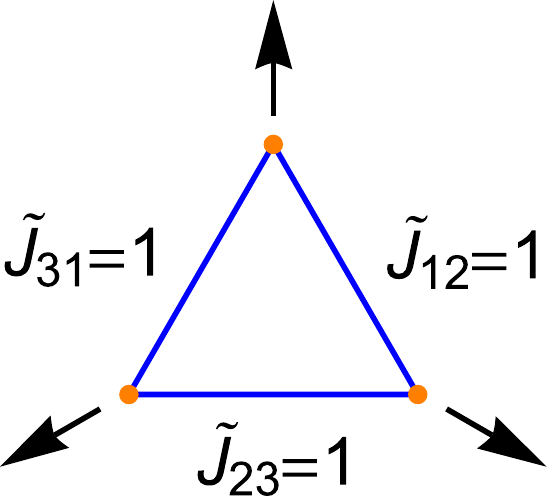}  & &
		\includegraphics[width=2.1cm]{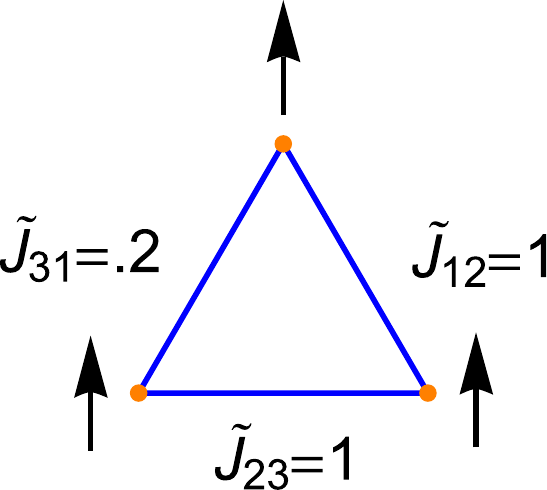}   
		\\		
		(a) &  & (b)
		\\
		&&
		\\
		\includegraphics[width=4cm]{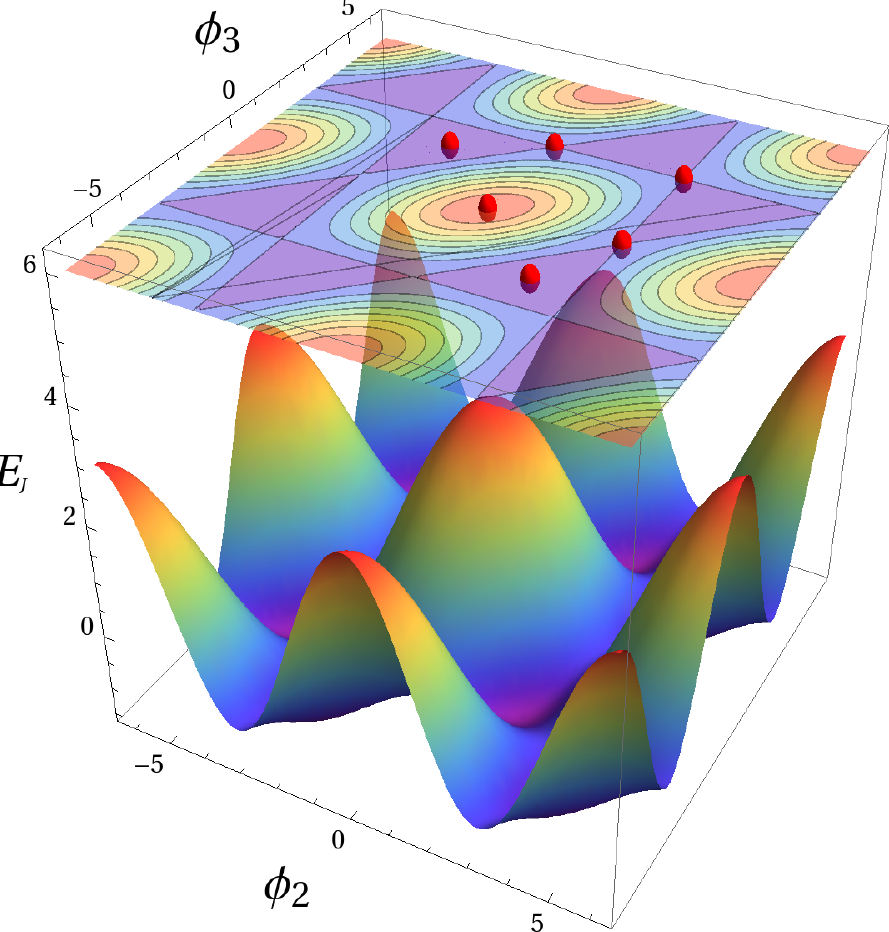}  & &
		\includegraphics[width=4cm]{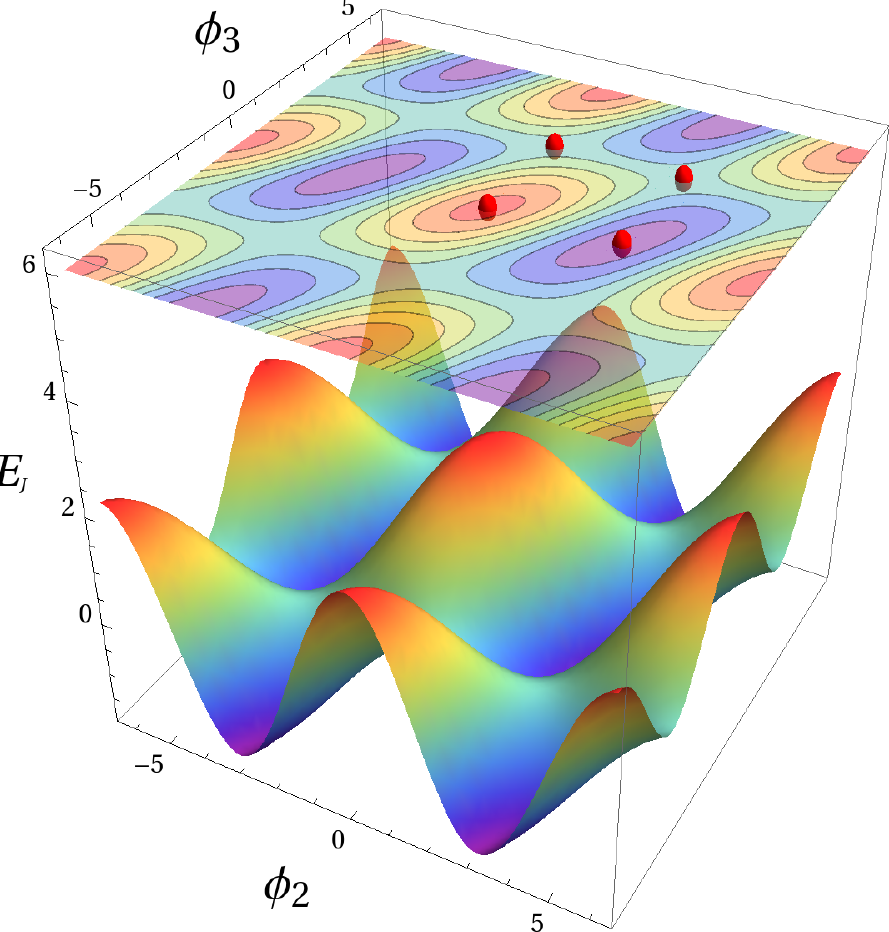}   
		\\
		(c) &  & (d)
		\end{array}$
	\end{center}
    \caption{\label{fig:energyplots} (c) and (d): Plots of the  Josephson energy $E_J$ as a function of the phases $\phi_2$ and $\phi_3$ for respectively: (a) a frustrated array of three classical spins ; (b) a non-frustrated array. The Josephson couplings are indicated in (a) and (b). Two additional solutions of the minimization conditions (red spheres in the contour plot) are present in the frustrated case.}
\end{figure}
%%%%%%%%%%%%%%%%%%%%%
%      end          %
%%%%%%%%%%%%%%%%%%%%%

The phases associated with superconducting parameters, $\Psi_i= \vert \Psi_i\vert e^ {i \phi_i}$,  only affect the interband contributions for the ground state energy. The minimization of this energy with respect to the phases $\phi_{i}$ gives 
\begin{eqnarray}
     \frac{\partial E}{\partial \phi_{i}} 
     = 
     0, \, \forall i & \Rightarrow & \label{eq:minimization} \\
     & &  \hspace{-1cm} \sum_j V_{i,j} \left\vert \Psi_{j} \right\vert
     \sin (\phi_{j}-\phi_{i})
     =
     0, \, \forall i. \nonumber 
\end{eqnarray}

These equations motivate a simple analogy with a system of $n$ two-component classical spins governed by the Heisenberg Hamiltonian $H=\sum_{i>j} J_{ij} \vec{S}_i \cdot \vec{S}_j$. The two-component classical spin is written as $\vec{S}_i= S_i (\cos \phi_i, \sin \phi_i)$. The energy of this system for a given set of magnitudes $\{S_i \}$ and angles $\{\phi_i \}$ is $E_J(\{ \phi_i \})=\sum_{i>j} J_{ij} S_i S_j \cos (\phi_i-\phi_j)=\sum_{i>j} \tilde{J}_{ij} \cos (\phi_i-\phi_j)$ where $\tilde{J}_{ij}=J_{ij} S_i S_j$. The last expression of this  energy  can also be interpreted as the energy of a  triangular circuit of Josephson junctions with couplings $\tilde{J}_{ij}$. The classical ground state configuration  is obtained from the minimization of this energy with respect to the angles $\phi_i$ which leads to the condition $\sum_{j} \tilde{J}_{ij} \sin (\phi_i-\phi_j)=0$ for each $i$, analogous to Eq.~\ref{eq:minimization} with the correspondence $\tilde{J}_{ij} \rightarrow -2 V_{ij} \vert \psi_i \vert \vert \psi_j \vert$. 
Note that if a set of angles $\{\phi_i \}$ is a solution of the previous set of equations, then the set  $\{-\phi_i \}$ also is. This implies that  non-frustrated solutions must have  $\phi_i-\phi_j=0,\pm \pi$. Other values correspond to frustrated configurations (degenerate ground states which are chirally different). 

We will now restrict our study to three bands ($n=3$) but the arguments are easily generalized to any $n$.
We impose for a matter of convenience that $\phi_{1}=0$ (with no loss of generality). Solving Eqs.~\ref{eq:minimization} we get several solutions corresponding to extreme or saddle points of the interband energy contribution, the non-frustrated solutions being  $(\phi_1,\phi_2,\phi_3)=(0,0,0)$, $(0,\pi,0)$, $(0,0,\pi)$, and $(0,\pi,\pi)$ and the frustrated solutions being
\begin{equation}
	(\phi_1,\phi_2,\phi_3)=\pm \left[0,\cos^{-1} (\alpha^-), -\text{sgn} \left(\dfrac{a}{b}\right) \cos^{-1}(\alpha^+) \right],
\end{equation}
where
\begin{equation}
	\alpha^{\pm}=\frac{\pm a^2 \mp b^2-a^2 b^2}{2 a^2 b},
\end{equation}
and $a=\tilde{J}_{12}/\tilde{J}_{23}$ and $b=\tilde{J}_{31}/\tilde{J}_{23}$. These frustrated solutions exist only if  $\vert \alpha^{\pm} \vert \leq 1$.

%%%%%%%%%%%%%%%%%%%%%
%      figure       %
%%%%%%%%%%%%%%%%%%%%%
\begin{figure}
	\includegraphics[width=6.5cm]{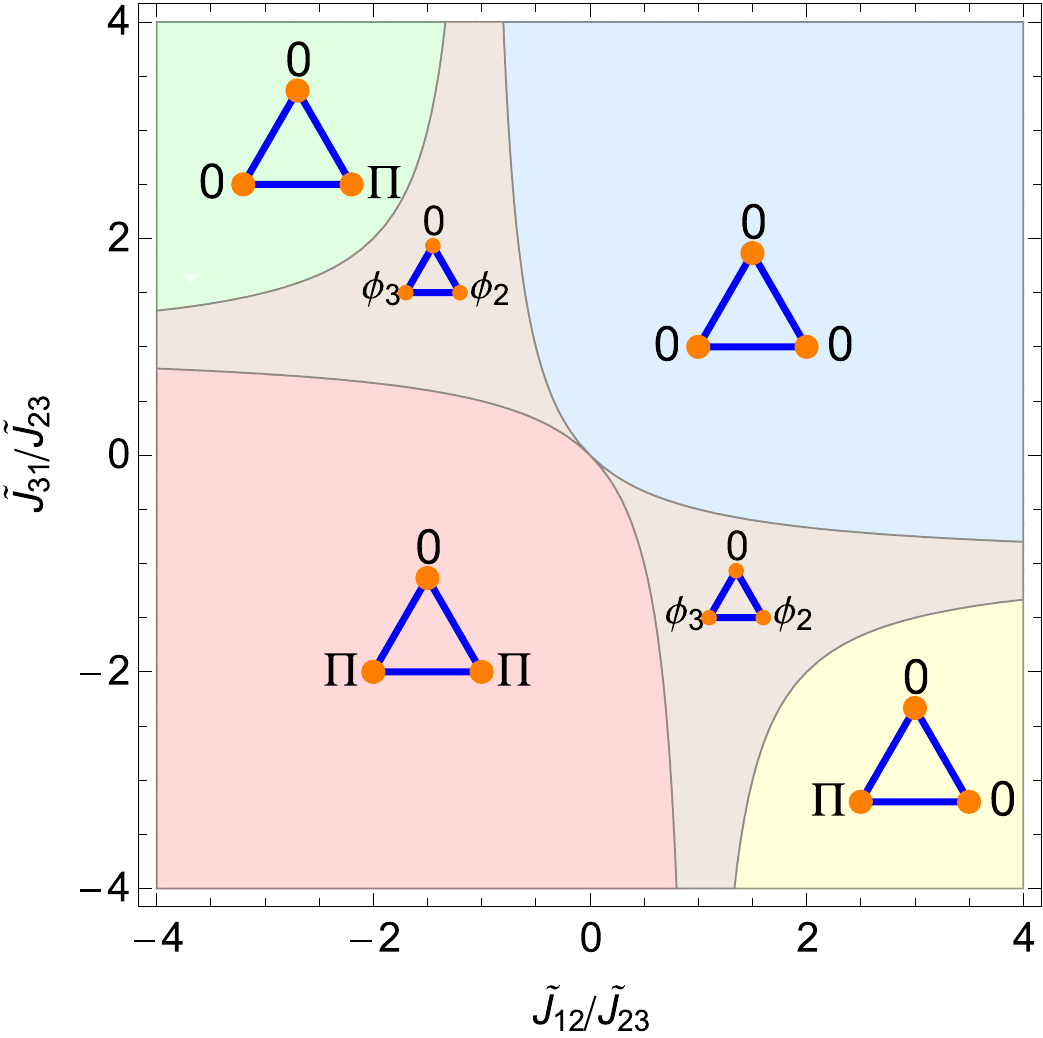}  
	\caption{The $\tilde{J}_{31}/\tilde{J}_{23}$ versus $\tilde{J}_{12}/\tilde{J}_{23}$ phase diagram of three classical spins  when one of the interactions $\tilde{J}_{ij}$ is negative. The non-frustrated regions have $(\phi_1,\phi_2,\phi_3)=(0,0,0)$, $(0,\pi,0)$, $(0,0,\pi)$, and $(0,\pi,\pi)$. The boundaries of the regions of frustration correspond to second-order phase transitions.}
	\label{fig:phasediagram}
\end{figure}
%%%%%%%%%%%%%%%%%%%%%
%      end          %
%%%%%%%%%%%%%%%%%%%%%
In Fig.~\ref{fig:energyplots}(a) and Fig.~\ref{fig:energyplots}(b) we show a frustrated array of three classical spins and a non-frustrated array, respectively.  In Fig.~\ref{fig:energyplots}(c) and Fig.~\ref{fig:energyplots}(d),  the respective plots of $E_J$ as a function of the phases $\phi_2$ and $\phi_3$ as well as the respective contour plots are displayed. The location of the solutions of the minimization equations  is indicated by the red spheres in the contour plots.  As $\tilde{J}_{31}$ is varied from 1 to .2, the position of the frustrated solutions  in Fig.\ref{fig:energyplots}(c) continuously converges to the saddle point $(\phi_2,\phi_3)=(0,\pi)$ [and to the equivalent location $(0,-\pi)$] and disappears when reaching this point [with $(0,\pi)$ becoming a local minimum]. This is a typical second-order phase transition.

In Fig.~\ref{fig:phasediagram}, the $\tilde{J}_{31}/\tilde{J}_{23}$ versus $\tilde{J}_{12}/\tilde{J}_{23}$ phase diagram of the three classical spins system  is displayed in the case when
one of the interband  $\tilde{J}_{ij}$ is negative. Note that in this case, the ratios of the  couplings  determine the signs of the couplings fully. The phase diagram of the three classical spins system when one of the interband $\tilde{J}_{ij}$ is positive can be easily obtained from the former using  the transformation $\phi_3 \rightarrow  \phi_3+\pi$ which leads to $\tilde{J}_{23} \rightarrow  -\tilde{J}_{23}$ and $\tilde{J}_{31} \rightarrow  -\tilde{J}_{31}$. For example, the symmetrically frustrated case $\tilde{J}_{12}=\tilde{J}_{23}=\tilde{J}_{31}=1$ has the frustrated solutions $(\phi_1,\phi_2,\phi_3)=\pm (0,2 \pi/3, -2 \pi/3)$.  These solutions using the previous transformation correspond to the frustrated case $\tilde{J}_{12}=1$ and $\tilde{J}_{23}=\tilde{J}_{31}=-1$ and $(\phi_1,\phi_2,\phi_3)=\mp (0,\pi/3, 2 \pi/3)$.
Recalling the relation with the multiband superconductor,  one has $\tilde{J}_{31}/\tilde{J}_{23} \rightarrow (V_{31}/V_{23}) \cdot (\vert\Psi_{1}\vert / \vert\Psi_{2}\vert)$ and $\tilde{J}_{12}/\tilde{J}_{23} \rightarrow (V_{12}/V_{23}) \cdot (\vert\Psi_{1}\vert / \vert\Psi_{3}\vert)$, so that the location of the multiband superconducting system in this phase diagram is determined not only by the interband pairings but also by the superconducting gaps. 

We have discussed in the previous paragraphs the minimization with respect to the superconducting phases. Given a certain phase configuration, the remaining minimization with respect to the absolute values of the superconducting order parameters gives the following result for $n$ bands,
\begin{equation}
        \Delta_{i\bm{k}} = \sum_{j \bm{k}'} 
        V_{\bm{kk}'}^{ij} \cos(\phi_{j}-\phi_{i}) u_{j\bm k'} v_{j\bm k'},
\end{equation}
so that the effect of the superconducting phase differences is the renormalization of the interband coupling.

%%%%%%%%%%%%%%%%%%%%%%%%%%%%%%%%%%%%%%%%%%%%%%%%%%%%%%%%%%%%%%%%%%%%%%%%
%         Finite temperature                                           %
%%%%%%%%%%%%%%%%%%%%%%%%%%%%%%%%%%%%%%%%%%%%%%%%%%%%%%%%%%%%%%%%%%%%%%%%
%\section{Finite temperature}
The same reasoning can be followed at finite  temperature. One must consider the free fermion entropy contribution for the free energy as well as  the non-zero occupation of quasiparticle states. One has
\begin{equation}
        \Psi_i =\sum_{\bm k} u_{i\bm k} v_{i\bm k}  e^{-i \phi_j}
        (1-2 f_{i\bm k}),
\end{equation} 
and the interband pairing term has the same expression as that of Eq.~\ref{eq:zeroTenergy}. Since the other terms in the free energy do not depend on the superconducting phases, the minimization with respect to the superconducting phases generates the same set of equations as for zero temperature (Eq.~\ref{eq:minimization}).
Minimizing the free energy with respect to the absolute values of the superconducting parameters \cite{Shimahara1994,Leggett1966,Dias2005}, one obtains a system of coupled gap equations
\begin{equation}
        \Delta_{i\bm{k}} = \sum_{j \bm{k}'} 
        V_{\bm{kk}'}^{ij} \cos(\phi_{j}-\phi_{i}) 
        u_{j\bm k'} v_{j\bm k'} (1-2 f_{i\bm k}),
        \label{eq:gapfiniteT}
\end{equation}
with $j=1,\dots , n $,
which, following the usual steps \cite{Shimahara1994,Dias2005}, can be written as 
\begin{equation}
         \Delta_i = \sum_{j} V^{ij}
         \cos(\phi_{j}-\phi_{i}) \int_0^{\omega_D} d\xi_j
         K_j(\xi_j,\Delta_j,T) \Delta_j ,
         \label{eq:gapequation1}
\end{equation}
with
\begin{equation}
         K_j(\xi,\Delta,T)={N_j(\xi) \over  E}
          \tanh {E \over 2t} ,
\end{equation}
where  $ E=\sqrt{\xi^2+\Delta^2}$, $\omega_D$ is the usual frequency cutoff, $N_j(\xi)$ is the density of states of the $j$ band and $t=k_B T$. We assume  equal constant density of states for all bands,  $N_j(\xi)=N_j(0)=N$, as a simplification. The differences in the density of states could also be absorbed in the couplings definition.

%%%%%%%%%%%%%%%%%%%%%
%      figure       %
%%%%%%%%%%%%%%%%%%%%%
\begin{figure}
\includegraphics[width=.9\linewidth]{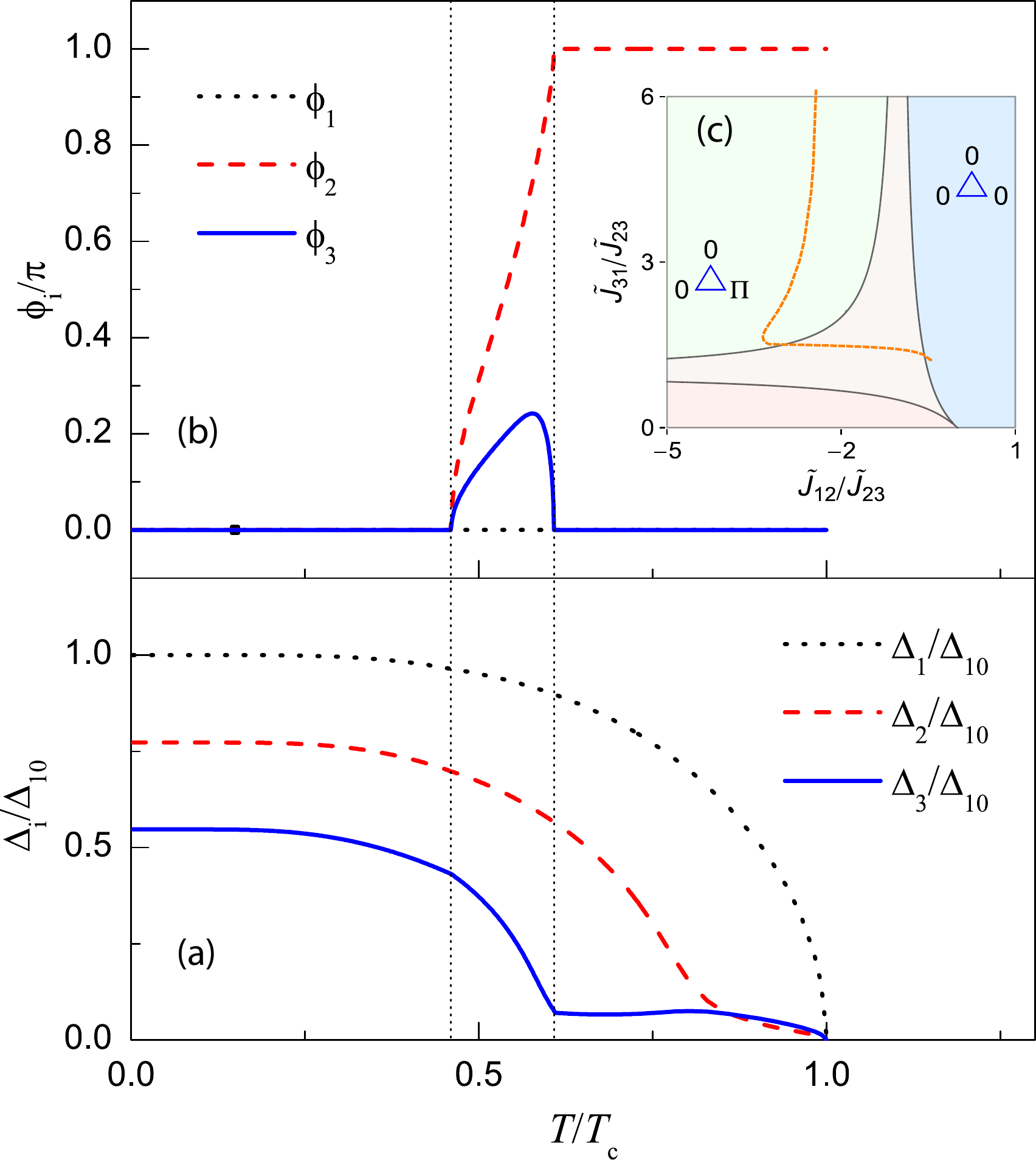} 
        \caption{\label{fig:deltasphasediagram} (a) Superconducting gap functions  and (b) superconducting phases as function of the normalized temperature  for a three-band superconductor with one repulsive and two attractive interband couplings. The vertical dotted lines delimit the temperature range where frustration occurs. Slopes changes are clearly observed in the lower superconducting gap at the boundaries of this interval indicating second-order phase transitions. (c) Path followed (dashed orange curve) in the $\tilde{J}_{31}/\tilde{J}_{23}$ versus $\tilde{J}_{12}/\tilde{J}_{23}$ phase diagram by the corresponding three classical spin system. Parameters: $V_{12}=-0.0045$, $V_{22}=.95$, $V_{23}=0.016$, $V_{31}=0.016$  and $V_{33}=0.85$ in units of $V_{11}$.
        }
\end{figure}
%%%%%%%%%%%%%%%%%%%%%
%      end          %
%%%%%%%%%%%%%%%%%%%%%

The finite temperature minimization process with respect to the superconducting phases is the same as that described for zero temperature (with the same analogy to the classical spin system), but now the couplings $\tilde{J}_{ij}$ are temperature dependent. Let us  consider the case where the interband couplings, $V_{ij}$, are finite, but much smaller than the intraband pairings, $V_{ii}$. The bands are indexed according to their uncoupled critical temperatures (critical temperature associated with each band when all interband couplings are zero), starting at the highest $T_c$ and ending at the lowest. 

In the uncoupled situation, the superconducting gaps display  sharp decreases at the respective critical temperatures with temperature regions where only $\Delta_3$ is zero (between $T_{c3}$ and $T_{c2}$) and  both $\Delta_3$ and $\Delta_2$ are zero (between $T_{c2}$ and $T_{c1}$). The modifications relatively to the uncoupled case, when the interband couplings $V_{ij}$ are small, are larger in these temperature ranges since the coupling of the gap equations implies that if $\Delta_1$ is non-zero at a local minima, then $\Delta_2$ and $\Delta_3$ are also non-zero. So in the non-frustrated situation one expects for the three-band superconducting gaps an analogous temperature dependence to that typical of a two-band superconductor (with an extra gap function) \cite{Suhl1959}. 

However, the coupling $\tilde{J}_{ij}$ depends strongly on the values of the superconducting parameters $\vert\Psi_{i}\vert$ and $\vert\Psi_{j}\vert$ which, for weak interband interactions, follow approximately the behavior of the uncoupled gap functions, i.e., undergo sharp decreases around the respective uncoupled critical temperatures. Therefore one should expect a fast increase in the absolute value of  $\tilde{J}_{31}/\tilde{J}_{23} \rightarrow (V_{31}/V_{23}) \cdot (\vert\Psi_{1}\vert / \vert\Psi_{2}\vert)$ around the uncoupled critical temperature $T_{c2}$ and  a similar increase of $\tilde{J}_{12}/\tilde{J}_{23} \rightarrow (V_{12}/V_{23}) \cdot (\vert\Psi_{1}\vert / \vert\Psi_{3}\vert)$ around the uncoupled critical temperature $T_{c3}$. Such variations may lead to crossings from a frustrated region of the phase diagram to a non-frustrated region or vice-versa. This is shown in Fig.~\ref{fig:deltasphasediagram}(c)  for a three-band superconductor which is initially non-frustrated and completely crosses the frustration region with increasing temperature. This system has weak (one repulsive and two attractive) interband couplings: $V_{12}=-0.0045$, $V_{22}=.95$, $V_{23}=0.016$, $V_{31}=0.016$  and $V_{33}=0.85$ in units of $V_{11}$.
The plot of the  superconducting gap functions as function of the normalized temperature for this system is shown in Fig.~\ref{fig:deltasphasediagram}(a). The associated  superconducting phases are displayed in  Fig.~\ref{fig:deltasphasediagram}(b) (only one of the frustrated solutions is shown, the other being symmetric).
Both graphs were obtained solving numerically the minimization conditions, Eqs.~\ref{eq:minimization} and \ref{eq:gapfiniteT}.
In Fig.~\ref{fig:deltasphasediagram}(a), slopes changes are observed in the lower superconducting gap which reflect a second-order phase transition from or to a frustrated phase configuration as shown in Fig.~\ref{fig:deltasphasediagram}(b). The vertical dotted lines indicate these transitions. Note that tiny slope changes also occur in the other superconducting gap curves. These changes are very small because the interband contribution to the respective gap function values is much smaller than the intraband contribution.
In  Fig.~\ref{fig:deltasphasediagram}(b), the superconducting phases do not reach the point $(\phi_1,\phi_2,\phi_3)=\mp (0,\pi/3, 2 \pi/3)$ corresponding to maximum frustration. This is because the $\vert \tilde{J}_{ij} \vert$ are never simultaneously equal to one as one increases temperature.

The transitions to or from frustrated configurations are more difficult to occur if the interband pair tunnellings is of the order of or stronger than the intraband pairing, since the ratios of the superconducting gaps in this case have little temperature dependence leading to short paths in the phase diagram of Fig.~\ref{fig:phasediagram}. This can be illustrated considering the particular case when $V_{ij}=\alpha^{i+j-2} V_{11}$, with $\vert \alpha \vert <1$, which has a non-frustrated solution  $\Delta_3 = \alpha \Delta_2= \alpha^2 \Delta_1$, which implies that the  ratios $\tilde{J}_{31}/\tilde{J}_{23}$ and  $\tilde{J}_{12}/\tilde{J}_{23}$ are constant and therefore the path in the phase diagram of Fig.~\ref{fig:phasediagram} becomes a single point.
One should emphasize that even with weak interband couplings, in the case described above of  one repulsive and two attractive interband couplings, if the  three-band superconductor is initially in the $(0,\pi, 0)$ non-frustrated region, no transitions will be observed with increasing temperature.

As far as we know, there are at the present no experimental reports of the temperature-induced phase transitions reported in this manuscript. One should recall that necessary conditions for the existence of such effects are: i) more than two bands participate in the superconducting state; ii) attractive intraband interactions; iii) weak interband pair tunnellings; iv) odd number of repulsive interband interactions. Furthermore, the second-order phase transitions to or from frustrated configurations are only clearly observed in temperature ranges where the interband contribution is dominant at least for one of the gap functions.
Among the several examples of superconductors where the possibility of  multiple bands contributing to the formation of a superconducting phase has been considered, the 
iron-based layered pnictides seem to be the more suitable candidates for the observation of these effects due to the complex band structure with more than two Fermi surfaces (hole and electron like).

%%%%%%%%%%%%%%%%%%%%%%%%%%%%%%%%%%%%%%%%%%%%%%%%%%%%%%%%%%%%%%%%%%%%%%%%%%
%                               Conclusion                               %
%%%%%%%%%%%%%%%%%%%%%%%%%%%%%%%%%%%%%%%%%%%%%%%%%%%%%%%%%%%%%%%%%%%%%%%%%%
%\section{Conclusion}
In conclusion, we have studied a three-band superconductor considering  frustration effects due to the existence of repulsive interband interactions. 
With increasing temperature and in the case of small interband couplings, one may have second-order phase transitions to or from frustrated configurations of the superconducting phases which lead to slope changes in the temperature dependence of the superconducting gaps.

\bibliography{frustrated}

\end{document}